\documentclass[onecolumn]{revtex4}
\pdfoutput=1
\usepackage{amsmath}    
\usepackage{graphicx}   
\usepackage{verbatim}   
\usepackage{color}      
\usepackage{subfigure}  
\usepackage{hyperref}   
\usepackage{subfigure}
\usepackage{float}
\begin{document}


\title{Entrainment of marginally stable excitation waves by spatially extended sub-threshold periodic forcing}

\author {Joseph M. Starobin, Vivek Varadarajan\\}
\mbox{}\\
\address{ Joint School of  Nanoscience and Nanotechnology, North Carolina Agricultural
and Technical State University, North Carolina 27401\\
 University of North Carolina at Greensboro, North Carolina, 27402\\
}

\begin{abstract}

We analyze the effects of spatially extended periodic forcing on the dynamics of one-dimensional excitation waves. Entrainment of unstable primary waves has been studied numerically for different amplitudes and frequencies of additional sub-threshold stimuli. We determined entrainment regimes under which excitation blocks were transformed into consistent 1:1 responses. These responses were spatially homogeneous and synchronized in the entire excitable medium.  Compared to primary pulses, pulses entrained by secondary stimulations were stable at considerably shorter periods which decreased at higher amplitudes and greater number of secondary stimuli. Our results suggest a practical methodology for stabilization of excitation in  reaction-diffusion media with regions of reduced excitability. 
\end{abstract}

\maketitle


Dynamics of excitation waves in reaction-diffusion media can be altered by spatio-temporal periodic forcing. Additional (secondary) periodic stimulations superimposed on primary forcing may alter the primary excitation waves and entrain (lock) them to the period of secondary stimuli.

Locking of primary waves to the period of secondary stimulations occurs at particular values of forcing periods and amplitudes. This resonant shift is characterized by Arnold tongues which determine different types of $M:N$ ($M\ge N,\;N\ge 1$) locking responses as a function of amplitude and frequency of external forcing  \cite{PRE2004,PRL2000}. 

It was found that phenomenon of locking manifests itself in different ways depending on spatio-temporal complexity of a particular reaction-diffusion system. For example, under the periodic forcing spatially uniform two-dimensional BZ reaction oscillations were transformed into standing wave type labyrinths of complex geometry \cite{PRE2004}. It was also demonstrated that one-dimensional Turing patterns can be modulated by spatio-temporal forcing in the form of traveling wave \cite{PRL2003}. 

Similar type of resonant behavior of Turing patterns was observed in experiments with photosensitive chemical reactions \cite{PRL2004}. 
It was also shown that the presence of large amplitude periodic forcing in a one-dimensional bistable reacion-diffusion medium resulted in bifurcating of originaly stable wavefront into two counter propagating wavefronts \cite{Europhy03}. In contrast, modulating the intensity and frequency of the periodic forcing controlled the trajectory and rotational frequency of two-dimensional spiral waves in the Oregonator model \cite{Europhy05}.  

One of the major parameters which affect reaction-diffusion dynamics is the amplitude of a threshold stimulus (excitation threshold) required to initiate an excitation wave.  While the outcome of over-threshold periodic stimulations has been discussed above in both experimental and theoritical settings \cite{PRE2004,PRL2000,PRL2003,PRL2004,Europhy03,Europhy05}, the response to periodic sub-threshold perturbabtions smaller than the excitation threshold has not been analyzed. 

In this paper, we study the entrainment of marginally stable excitation waves by spatially extended periodic sub-threshold forcing in the Chernyak-Starobin-Cohen reaction-diffusion medium \cite{Chernyak98}. 
\begin{equation}
\begin{array}{ccc}
\displaystyle \frac{\partial u}{\partial t}&=&\displaystyle\frac{\partial^2 u}{\partial x^2}-i(u,v)+P(x,t)+\sum_{i=1}^{n}S_i(x_i,t)\\
i(u,v)&=&\begin{cases}
         \lambda u & \mbox{for }u<v\\
         (u-1) & \mbox{for }u\ge v
         \end{cases}\\
\displaystyle \frac{\partial v}{\partial t}&=&\epsilon( \zeta u+v_r-v)\\
\end{array}
\end{equation}

Here $u$ and $v$ are dimensionless excitation and recovery functions, respectively. $\epsilon$ is a small parameter and $v_r$ is the excitation threshold. $\lambda$ and $\zeta$ control the rates of changes of excitation and recovery variables, respectively.

The system of Eqs. 1 was solved numerically in a one-dimensional cable of finite length using a second order explicit difference scheme with zero flux boundary conditions \cite{Richtmayer}. Spatial, $\Delta x$, and temporal, $\Delta t$, steps used in the numerical integration were equal to $0.23$ and $0.0072$, respectively. The cable length, $L$, was equal to $150\Delta x$. Primary forcing, $P(x,t)$, was a periodic function with an over-threshold amplitude $A_0$ and period $T_0$. Primary stimuli were applied near the left end of the cable between $x=2\Delta x$ and $x=15\Delta x$. As in some biological reaction-diffusion media \cite{Ravens} we used a simplified primary rate dependent threshold given by linear equation $v_r=\alpha-\beta T_0,\;\alpha>0,\;\beta>0$.

Secondary forcing $\left\{ S_i(x_i,t),\;i=1,2,...,n\right\}$ was delivered at $n$ equidistant  locations between $x=40\Delta x$ and $x=140\Delta x$. The secondary stimuli were simultaneously activated after the wavefront resulted from the first primary stimulation arrived at the end of the cable. The amplitude and period of secondary stimuli were equal to $A$ and $T$, respectively. 

Unless mentioned otherwise, values of parameters in all computations were $\epsilon=0.1,\;\lambda=0.4,\;\zeta=1.2,\;A_0=1.4,\;\alpha=0.31,\;\beta=0.0025\mbox{ and } n=6$. The duration of a pulse, $T_h$, was measured as the time interval between consecutive intersections of $u$ and $v$ near their rest and excited states, respectiely (Fig. 1a). The steady state value of $T_h$ was computed after 80 primary stimulation periods.

In the absence of secondary stimulation ($A=0$), at long periods $T_0$ the variable $v$ has enough time to reach its ground state $v_r$ before the next stimulus is applied. However, as $T_0$ approaches a critical limit, $T_{end}$, the system does not respond to every stimulus and exhibits unstable $M:N$ ($M>N$) excitation blocks which occur due to incomplete recovery of control variable $v$. Figure 1b compares phase portraits of the system at two values of $T_0$.  At $T_0=60$, $u$ closely follows its nullcline and $v$ almost completely recovers to $v_r=0.16$ as depicted by the intersection of the $u$-$v$ nullclines. However, at $T_0=T_{end}=30$, deviations of $u$ and $v$ from their nullclines are quite significant,  thereby $v$ recovers to a value which is much higher than the corresponding threshold $v_r=0.23$.
\begin{figure}[htp]
\subfigure[]{ \includegraphics[scale=0.5]{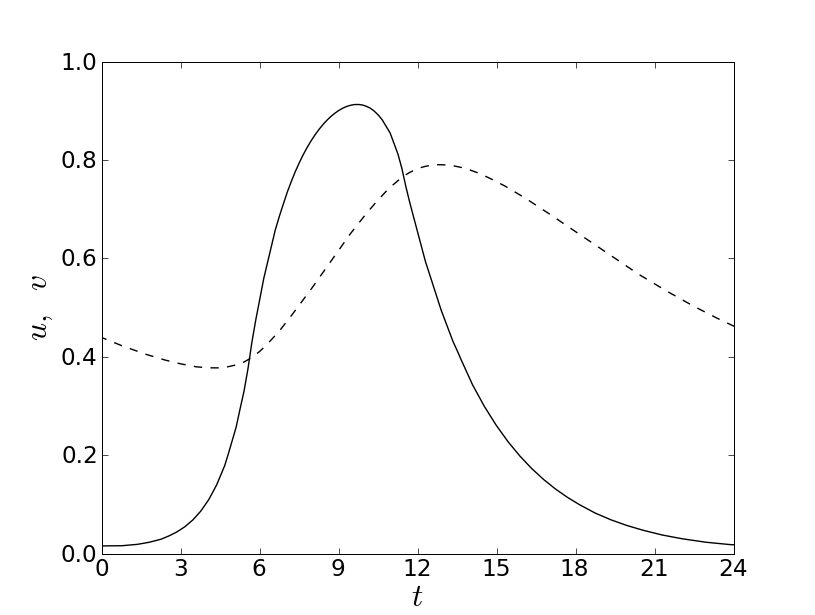}
}\\
\subfigure[]{\includegraphics[scale=0.5]{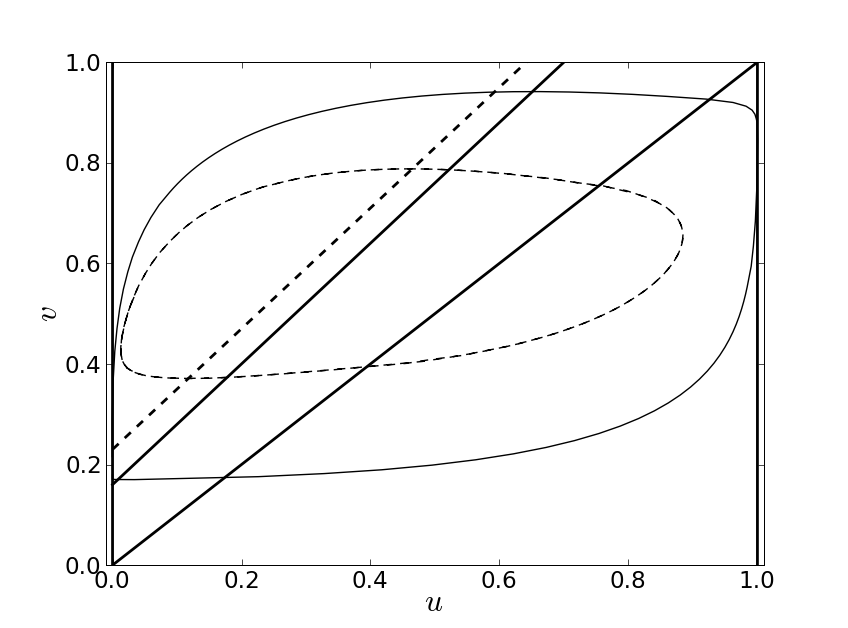}
}
\caption{Panel (a) shows excitation (solid line) and recovery (dashed line) variables at $x=L/2$ for the short steady state pulse. Left and right intersections between $u$ and $v$ mark the
begining and the end of the pulse, respectively. Panel (b) shows phase portraits of the system at the same location for $T_0 =T_{end}$ (thin-dashed contour) and $T_0=60$ (thin-solid con-
tour). Nullcine of $u$ is given by thick N-shaped line. Dashed and solid lines with intercepts at $v_r = 0.23$ and $v_r = 0.16$ are the v nullclines for $T_0 = T_{end}$ and $T_0 = 60$, respectively. }
\end{figure}

Analysis of pulse duration $T_h$ as a function of secondary frequency $F=T^{-1}$ and forcing amplitude $A\ge 0$ revealed a variety of entrainment regimes (Fig. 2). We found that the system did not respond to secondary stimuli at amplitudes which were smaller than critical values depicted by the curve with circular markers. For the amplitudes above this curve we observed intermediate $M:M$ responses with $M$ greater than one. For even greater amplitudes, above the upper curve with square markers, the system locked to secondary stimuli with consistent 1:1 responses. It should be noticed, that such locking occured in a wide range of frequencies of secondary stimulations at sub-threshold amplitudes which were 80$\%$ lower than amplitudes of primary stimuli. 
\begin{figure}[htp]
\includegraphics[scale=0.41]{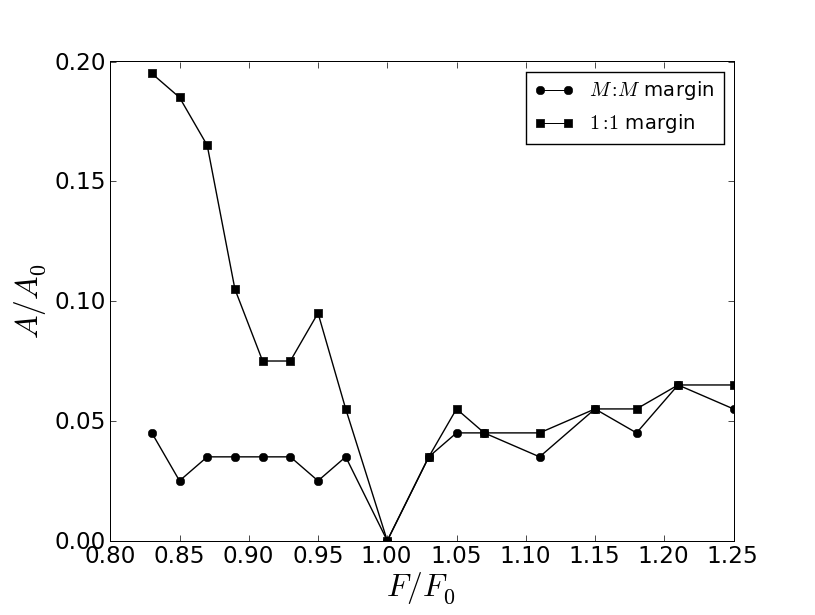}
\caption{Locking margins for different frequencies and amplitudes of secondary stimuli for $T_0=30, \;A_{0}=1.4$. $x=\displaystyle\frac{L}{2}$}
\end{figure}
 
\begin{figure}[htp]
\subfigure[]{ \includegraphics[scale=0.35]{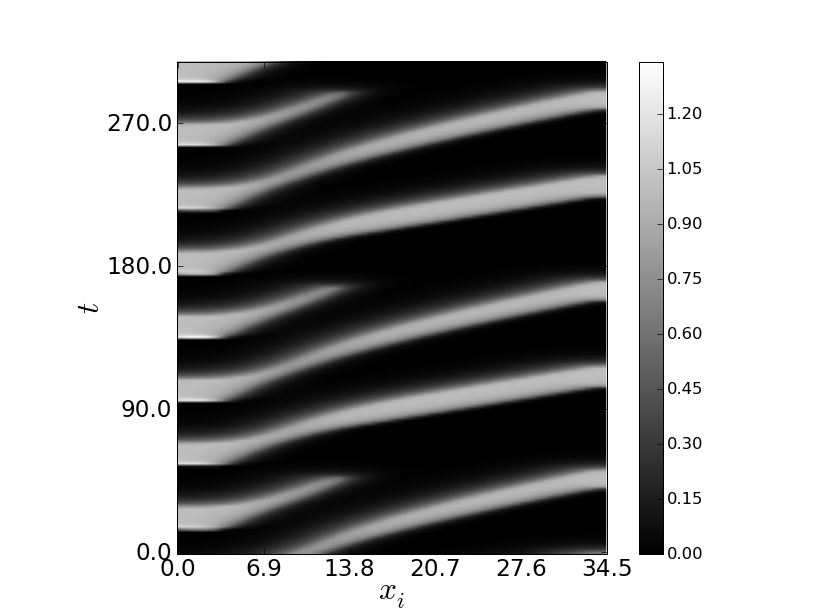}
}\\
\subfigure[]{\includegraphics[scale=0.35]{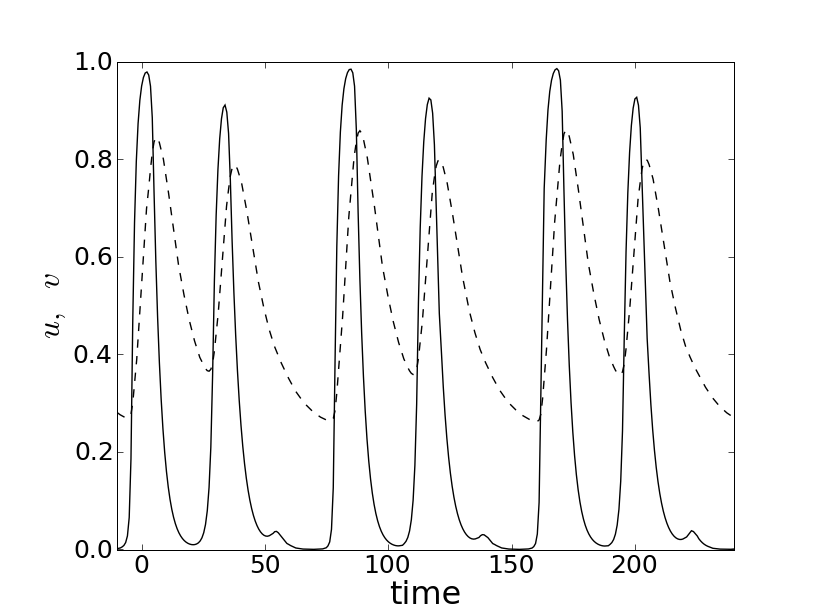}
}
\caption{Panel (a) shows the dynamics of $u$ as a function of time, $t$, and spatial location, $x_i=i\;\Delta x,\;\left\{i=0,1,...,150\right\}$. Panel (b) demonstrates temporal dynamics of $u$ (solid line) and $v$ (dashed line) for three consecutive cycles at $T_0=28,\;A=0$. $x=\displaystyle\frac{L}{2}$ }
\end{figure}
Spatio-temporal contours of $u$ shown in Fig. 3a demonstrate expected unstable responses to primary stimulation at $T_0<T_{end}$. Temporal dynamics of $u$ and $v$, as well as $u$-contours, show 3:2 excitation blocks (Fig. 3b). However, in the presence of secondary stimulations such unstable responses can be entrained and stabilized by secondary driving even at $T_0<T_{end}$. Indeed, Fig. 4a demonstrates that 3:2 blocks can be transformed into stable 1:1 responses which evolve homogeneously in the entire cable except short segments located near the site of primary stimulation. Formation of these fully synchronized responses are preceded by very short ($\sim 0.02T_{0}$) transient periods  during which standing wave type oscillations of $u$ rapidly saturate at constant excitation levels (Fig. 4b).

When compared with primary forcing alone ($A=0$), secondary sub-threshold stimuli facilitate development of stable pulses shorter than those at $T_{end}=30$.  Fig. 5a shows that secondary stimuli of higher amplitudes sustain shorter entrained pulses at progressively lower stimulation periods. The trend is slightly augmented for higher coefficients $\beta$ and lower amplitudes of secondary stimuli (Fig. 5b).

Stabilization of the system due to secondary driving can be also achieved using greater number of secondary stimulation sources. Correspondingly, Fig. 6 demonstrates that due to initial two-fold increase of the number of secondary sources the region of stability is significantly extended towards shorter values of $T_{end}$. Further increase of the number of sources saturates these changes at higher rates for smaller values of the coefficient $\beta$. Under these conditions excitation thresholds are lower, pulse wavefronts are narrower and, therefore, stabilization of dynamics is achieved at smaller values of $T_{end}$.
\begin{figure}[!htp]
\subfigure[]{
\includegraphics[scale=0.35]{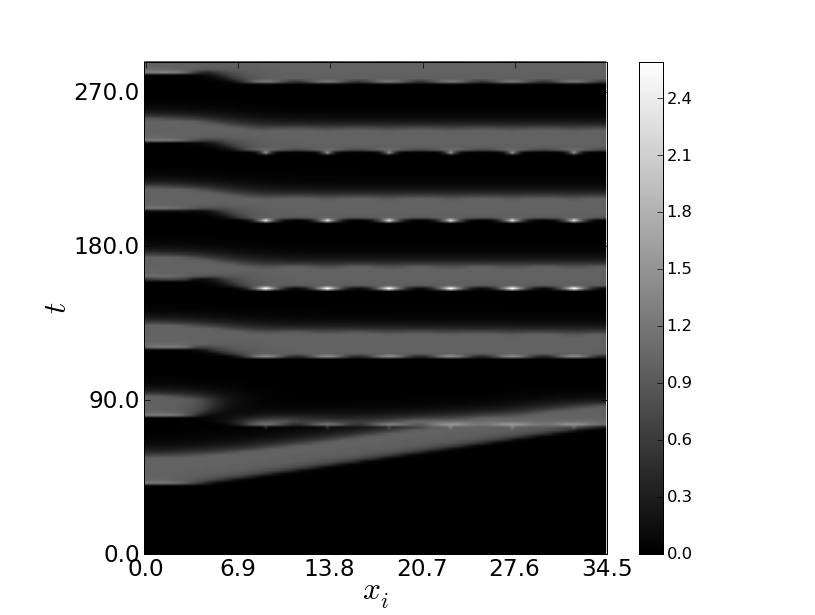}}
\subfigure[]{
\includegraphics[scale=0.35]{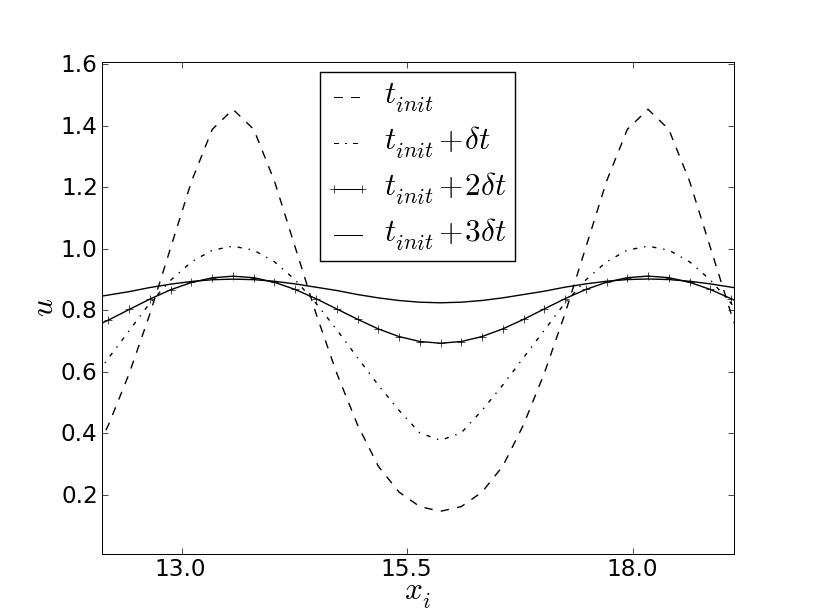}}
\caption{Panel (a) shows spatio-temporal band type synchronized patterns of $u$ ($T_0=T=28$) for $A=0.1A_0$. Panel (b) depicts temporal evolution of a spatial profile of $u$ at three equidistant moments of time, starting from $t_{init}=136$, $\delta t=0.36$. }
\end{figure}
\begin{figure}[htp]
\subfigure[]{
\includegraphics[scale=0.35]{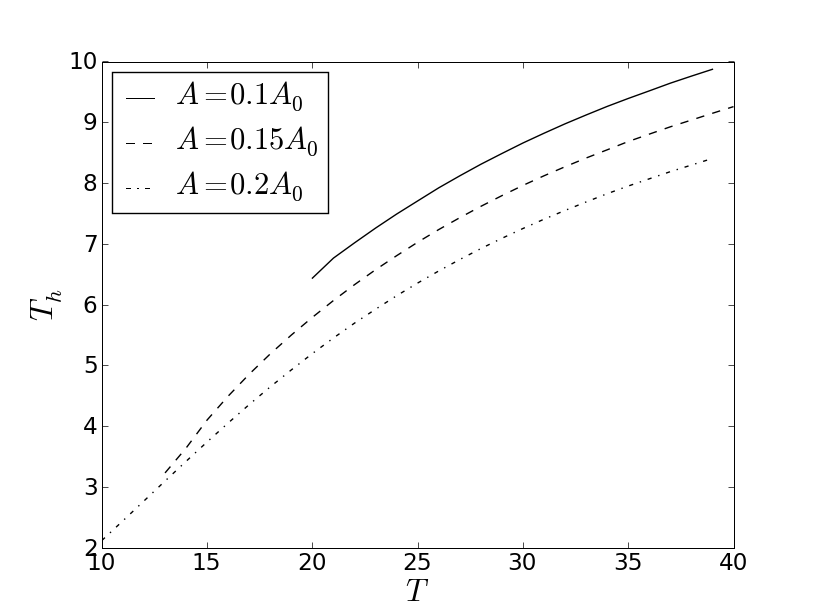}}
\subfigure[]{
\includegraphics[scale=0.35]{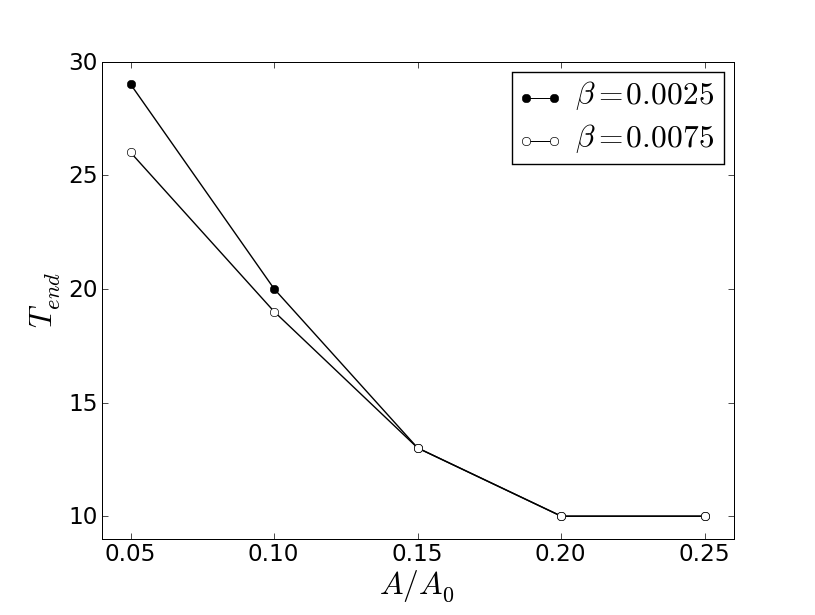}}
\caption{Panel (a) shows dependence of $T_h$ on $T$ for different amplitudes of secondary stimulli, $T_0=T$. For each amplitude ratio, the lowest value of $T$ corresponds to $T_{end}$. Panel (b) depicts dependence of $T_{end}$ on $\displaystyle\frac{A}{A_0}$ for two values of $\beta$. $\alpha=0.31$ and $x=\displaystyle\frac{L}{2}$.}
\end{figure}
\begin{figure}[htp]
\includegraphics[scale=0.38]{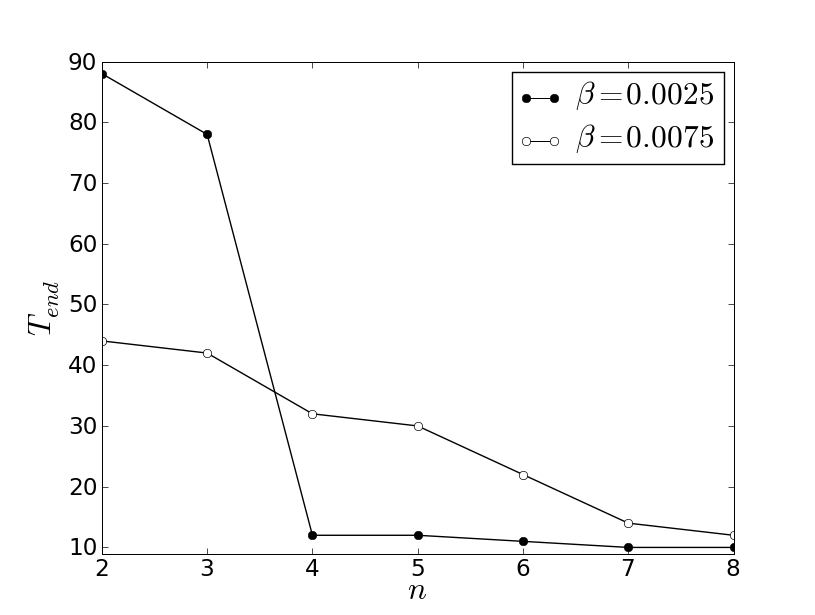}
\caption{Dependencies of $T_{end}$ on $n$ for different $\beta$. $\alpha=0.67$ and $x=\displaystyle\frac{L}{2}$.}
\end{figure}

In summary, we demonstrated that additional sub-threshold driving stimuli can entrain otherwise unstable primary reaction-diffusion waves and transform $M:N$ excitation blocks into stable 1:1 spatially homogeneous responses synchronized  in the entire cable. Compared to pulses resulted from primary forcing alone, pulses entrained by secondary stimulations were stable at considerably shorter periods. These periods decreased at higher amplitudes and greater number of secondary stimuli. Our results outline the possibility of entrainment of reaction-diffusion waves by sub-threshold additional driving and may be applied for stabilization of excitation in  reaction-diffusion systems with zones of reduced excitability \cite{Arava01}. 

\section*{Acknowledgements}
We would like to thank Vladimir Polotski and Shyam Aravamudhan for useful discussions and continuous interest to our work. We also thank Alan Covell for editorial comments.

 \bibliography{PRL_bib}
\end{document}